# Trade-offs in the Design of Blockchain of Finite-Lifetime Blocks for Edge-IoT Applications

Shravan Garlapati, IEEE Senior Member, Email: gshra09@vt.edu, ORCID : 0000-0001-6203-3770

*Abstract*— Unlike cryptocurrency transactions in bitcoin that are stored indefinitely, the data of certain applications like IoT have finite-lifetime. In this context, one of the recent research works proposed LiTiChain - a new architecture for Blockchain of finite-lifetime blocks with applications to Edge-IoT. The novelty of LiTiChain lies in ensuring the connectivity of the chain even after the expired blocks are deleted from the chain. To provide the same level of security as conventional blockchain, in LiTiChain, some blocks are stored longer than their lifetime, which incurs additional storage cost. This paper presents two new blockchain architectures i.e. *p-LiTiChain* and *s-LiTiChain* that are variants of LiTiChain. The proposed architectures offer a degree of freedom in the design of blockchain of finite-lifetime blocks in terms of a tradeoff between storage cost, security and computational cost. With extensive simulations and analysis, it is demonstrated that the proposed architectures have the potential to decrease the additional storage cost incurred by LiTiChain to zero and improve security at the expense of computational cost.

I. INTRODUCTION

The Internet of Things (IoT) industry continues to grow rapidly, and it is forecasted that, by 2030, at least 50 billion devices will have internet connectivity [1]. IoT devices and sensors generate, process and exchange huge amounts of safety-critical and privacy-sensitive data and hence the security of these devices and data are a major concern. Considering the explosive number of devices, it is not a simple task to address issues in security, privacy and data integrity. To be specific, many of the state-of-the art security solutions are centralized and they may not fit well for IoT due to the massive scale of the generated data, single point of failure and many-to-one nature of the traffic. To overcome these issues, contrary to the traditional centralized approaches, numerous recent research works proposed the use of decentralized approaches such as Blockchain [2] [3] [4] [5].

Blockchain is a peer-to-peer (P2P) distributed system that offers improved security and privacy of data [6]. In the last decade, blockchain technology has gained a lot of attention as it is widely used in cryptocurrencies such as bitcoin, Ethereum, ripple etc. In recent years, efforts are underway to adopt blockchain technology in different sectors such as financial services, supply chain, healthcare, IoT etc. [7] [8] [9]. This paper considers a scalable blockchain architecture targeted at IoT applications in edge computing environment. As the number of connected IoT devices continue to explode exponentially, data storage and processing pose serious scalability problems to centralized cloud architectures. The *Edge computing* - a distributed computing paradigm alleviates these problems by bringing computation, data processing and storage closer to IoT devices to improve response times and save bandwidth [10] [11] [12]. In Edge computing, IoT data can be pre-processed at edge to compress and summarize the data sent to central cloud. This reduces the data that must be moved, and the distance data must travel, resulting in lower latency and reduced transmission costs. Edge computing offers middle-layer services between cloud and IoT device layer i.e. it not only handles data processing tasks for cloud but also takes care of computational offloading of resource constrained IoT devices for real-time applications such as audio and video recognition, autonomous cars, smart cities, industry 4.0 and home automation systems etc. [13] [14]. Thus, Edge computing has the potential to reduce computational load and power consumption of IoT devices [15] [16] [17].

In this context, few research works proposed *Edge-IoT* systems where edge-servers form a distributed network, employ blockchain to manage the allocation of edge resources to IoT devices, support storing and sharing of IoT data. In [18], Edge-chain, a credit-based resource management system employed blockchain to allocate edge server resource pool to IoT devices, manage all the IoT activities and transactions for secure data logging and auditing. Also, [19] proposed a cognitive Edge-IoT framework that hosts, and processes offloaded geo-tagged multimedia payload and transactions from IoT nodes and stores results in a blockchain and decentralized cloud repositories to support secure and privacy-oriented sharing economy services in a smart city. A decentralized storage, access control, data management and sharing system employing blockchain is presented in [20] to manage time-series IoT data at the edge. Thus, many research works applied blockchain for Edge-IoT applications, but except [21], none of them focused on *storage scalability*, a major challenge for *Edge-IoT*.

In Bitcoin Network (BCN), as of today, more than 250 GB of storage is required to store full blockchain. The throughput of the bitcoin is around 10 transactions per sec. On the contrary, considering the massive scale, transaction rate of IoT devices can be significantly higher compared to monetary transactions of bitcoin. Hence, the storage capacity needs of *Edge-IoT* can also be significantly higher compared to bitcoin and it is possible that the edge servers can ultimately run out of space to store the full chain. To address this issue, a recent research work proposed LiTiChain - a scalable and lightweight blockchain architecture [21]. In cryptocurrency systems, monetary transactions and blocks are stored permanently on the blockchain. On the other hand, IoT data has finite-lifetime and hence expired transactions and blocks can be deleted from the blockchain. LiTiChain exploited this idea and proposed a novel blockchain architecture to minimize the storage requirements. The disadvantage of LiTiChain is, instead of deleting the blocks immediately upon expiration, it is possible that some blocks are retained longer to validate remaining blocks, which results in *additional storage*

*cost*. To address this issue, two new architectures, *p-LiTiChain* and *s-LiTiChain*, which are variants of LiTiChain are proposed in this paper. The proposed architectures aim at reducing the *additional storage cost* incurred by LiTiChain and improving security of the chain. In this process, *p-LiTiChain* and *s-LiTiChain* provide a trade-off in the blockchain design (section IV) in terms of storage cost, security and computational cost.

The remainder of the paper is organized as follows: section II describes the preliminaries that lay the foundation to understand the subsequent sections. LiTiChain architecture is described briefly in section III. The architectures of *p-LiTiChain* and *s-LiTiChain,* along with the trade-off in blockchain design in terms of storage cost, security and computational cost are discussed in section IV. Section V presents the simulation results and analysis. Section VI concludes the paper.

## II. PRELIMINARIES

This section describes the basic concepts of blockchain that are derived from the Bitcoin [6] [21]. Blockchain can be defined as a specific type of distributed ledger that records transactions between two parties in an efficient and secure manner. In simple terms, blockchain is a growing list of records called as *blocks*, which are linked using cryptography [22]. As the blocks are connected using cryptographic hash, a block in a blockchain cannot be easily altered and hence the data is resistant to modification. In order to operate as a distributed ledger, a blockchain based system is typically managed by a P2P network. For example: in *Edge-IoT* systems, edge-servers form a P2P network. In BCN, certain users with large computational resources known as *miners,* participate in mining process. Every miner node independently aggregates transactions into a new block and when the new block is full, it is appended to blockchain by mining process i.e. miners solve a cryptographic puzzle with a certain difficulty known as *Proof of Work (PoW)*. A miner node that first solves the *PoW* appends the mined block to the blockchain and broadcasts the solution to the network. All miners in the network validate the solution, accept the updated blockchain and re-broadcasts the solution. The miner node that first solved the *PoW* is rewarded with bitcoins. The details of *Blockheader, Blockheight, and Nonce Computation* are explained below.

### A. Structure of a Block

As per [23], a block consists of header, metadata and long list of transactions. The Blockheader consists of three sets of metadata. First, *previous blockhash*, a pointer to the previous block. The second set, *timestamp, Nonce* and *difficulty* are related to the mining process. The third part is the *merkle tree root*. A block is identified by two identifiers, they are: *Blockhash* and *Blockheight*. *Blockhash* is the primary identifier and it is obtained by hashing the blockheader twice using the SHA256 algorithm. Another way to identify a block is by its position in the blockchain, known as *blockheight*. The first block, known as *Genesis block,* is at a blockheight of zero.

### B. Nonce Computation

*PoW* takes *previous blockhash, merkle root and timestamp* as input and computes hash using SHA256. The computational challenge involves finding a *Nonce* that results in an output hash that has certain number of leading bits as zero. The difficulty level in *Nonce computation* i.e. the number of leading bits as zeros in output hash is given by the *Difficulty Target* parameter specified in the blockheader. In bitcoin, for every new block, depending on the *Difficulty Target,* miners may test *billions or trillions* of *Nonce* before the requirements are met. The block is valid only if the miner succeeds in finding a *Nonce* that meets the target. The miner node broadcasts the valid block to the neighbors, which is further propagated to the BCN.

## III. LiTiCHAIN

As mentioned earlier, LiTiChain proposed a scalable and lightweight architecture for blockchain of finite-lifetime blocks targeted at edge-IoT applications. In LiTiChain, *lifetime* of a block is defined as the difference between creation and expiration times of the block. As in conventional blockchain, if the finite-lifetime blocks are chained in the order of their arrival times, when an expired block is deleted, it is possible that the blockchain can be disconnected. As shown in Fig. 1, when block $b_2$ is deleted at the end of its expiration time, it results in disconnected blockchain. Hence, in order to ensure chain connectivity even after the expired blocks are deleted, LiTiChain proposed a graph structure based on the expiration time of the blocks. Construction of Expiration time Ordering Graph (EOG) is as shown in Fig. 2. According to [21], the procedure to construct EOG is as follows: let us assume that the block created at the $i^{th}$ time instant is denoted by $b_i$ for $i = 1, 2, ....$ Also, let $t_i$ and $e_i$ respectively denote the creation and expiration time of a block $b_i$.

a) If there exist a set of blocks whose expiration time is later than the new block $b_i$. In this set, block $b_i$ is connected to the block with the earliest expiration time by a directed edge. Expiration time of the *Genesis block G* is infinity.

b) If expiration time of the new block $b_i$ is later than all the existing blocks, $b_i$ is connected to $G$ with a directed edge from $b_i$ to $G$.

As shown in Fig. 2, for every node*,* expiration time of the parent node is later than the child node. Hence, at all times, EOG remains a connected graph. But, the problem with EOG is, *blockheight* i.e. the distance measured from a block to the *Genesis block* can be short. In Fig. 2, blockheight of $b_5$ is 2 but in conventional blockchain the blockheight of $b_5$ would be 5. Additionally, as the expired blocks are deleted in EOG, unlike conventional blockchain, size of the chain may not grow, which results in shallow branches. As per the *longest* chain rule in the bitcoin blockchain i.e. longer the chain the harder it is for the attacker to undo the chain. In other words, longer chains are assumed to be more secure and hence are preferred [6] [21]. Therefore, to overcome the shallow EOG-based chains, another graph based on arrival time of blocks known as Arrival Ordering Graph (AOG) is coupled with the EOG to form LiTiChain*,* it is as shown in Fig. 3. In conventional blockchain, a block $b_i$ is connected to its previous block $b_{i-1}$ using *previousblockhash*. On the other hand, in LiTiChain, as shown in Fig. 4, every block is connected to two blocks - a parent block $b_{i^*}$ *via* EOG and previous block $b_{i-1}$ *via* AOG. In other words, hash $h_i$ of block $b_i$ depends on both *previous blockhash* $h_{i-1}$ and *parent blockhash* $h_{i^*}$. Therefore, as shown in Fig. 3*,* with the addition of AOG edges, *blockheight* of $b_6$ is increased from 4 to 6, which offers improved security as per the longest chain rule.

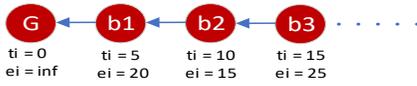

Fig. 1. Blockchain based on linear order of arrival times. Genesis block (G), Creation time ($t_i$) and Expiration time ($e_i$).

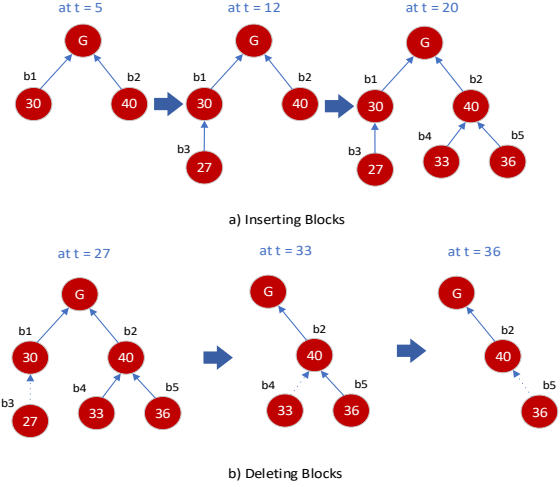

Fig. 2. Construction of expiration time ordering graph

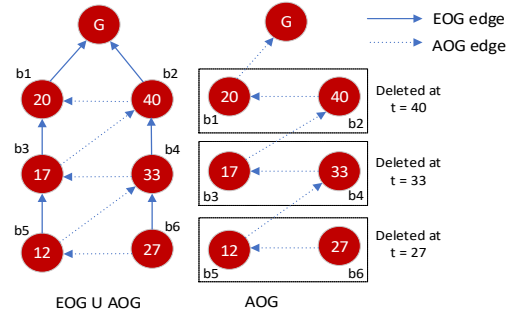

Fig. 3. Illustration of LiTiChain structure

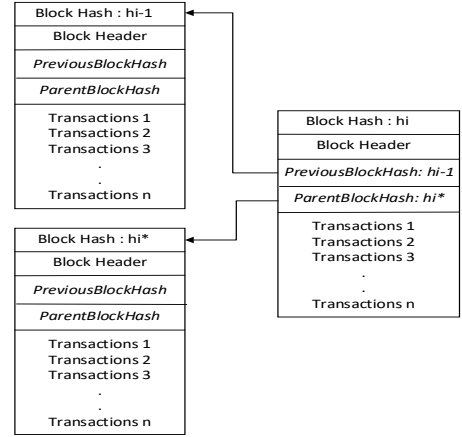

Fig. 4. Block Hash in LiTiChain

In Fig. 3, block $b_1$ should be ideally deleted at time instance 20. As there exists an AOG edge from $b_2$ to $b_1$ i.e. hash of $b_2$ depends on $b_1$, to validate $b_2$ i.e. verify that block $b_2$ is not corrupted by an attacker, $b_1$ needs to be stored until the expiration time of $b_2$ i.e. until time instance 40. This results in extra storage costs, known as *retention cost*. This is the disadvantage of LiTiChain. Section IV presents two new architectures that aim at reducing the *retention cost*.

## IV. PROPOSED SCHEMES

As mentioned earlier, as per the *longest chain rule*, from security perspective, higher values are preferred for the *blockheight*. But, in LiTiChain, higher values of *blockheight* results in higher retention cost. Let $K$ denote the *blockheight* threshold. In LiTiChain, if the blockheight of the parent of a newly added block $b_i$ is less than or equal to $K$, then an AOG edge is added between block $b_i$ and $b_{i-1}$, which incurs retention cost. On the other hand, if the blockheight of the parent is greater than $K$, AOG edge is not added and the retention cost of block $b_{i-1}$ is zero. As shown in Fig. 5, in LiTiChain, as the value of $K$ varies as 1, 2 and 4, overall retention cost increases to 20, 36 and 51. To reduce the retention cost, two new architectures that are variants of LiTiChain are presented in this paper. They are: *p-LiTiChain* and *s-LiTiChain*.

If an attacker intends to corrupt as many transactions as possible in a blockchain, he has to undo the total blockchain i.e. compute new hash for all the blocks. In this process, majority of the attacker's time and resources are spent in *Nonce computation* (Sec. II) i.e. finding a new *Nonce* for each block in the chain. Therefore, *if the number of AOG edges reduces while the height of the blocks increases or remains same as in LiTiChain, overall retention cost of the chain can be reduced without lowering attacker's difficulty in undoing the chain i.e. security of the chain is maintained at the same level.* Both *p-LiTiChain* and *s-LiTiChain* architectures aim at this.

As discussed in section II, in *public blockchain* systems like bitcoin, for every mined block, miners are rewarded with bitcoins. To reduce the number of bitcoins spent as a reward, BCN aims at reducing the number of mined blocks and hence the blocksize in bitcoin is generally higher. Unlike bitcoin, Edge-IoT employs *private blockchain*. Hence, miners are not rewarded for solving PoW. Therefore, if edge-servers have enough *Nonce computation* resources; it is ok to reduce the blocksize to increase the number of mined blocks, which increases the height of blocks in a blockchain. Hence, *when LiTiChain branches are shallower, blocksize can be reduced to increase the number of mined blocks added to the chain to increase the depth of LiTiChain branches*. As depth increases i.e. as the blockheight of the newly added blocks is greater than threshold $K$, blocksize can be increased i.e. the number of mined blocks can be reduced to reduce the number of *Nonce computations*. The proposed schemes exploit this principle to reduce the number of AOG edges and minimize the retention cost. For the same number of transactions to be processed, the reduction in blocksize increases the number of mined blocks and the number of *Nonce computations*, which leads to higher computational cost. So, there exists a trade-off between retention cost i.e. additional storage cost, security i.e. height of the blocks in the chain and *Nonce computational* cost. *The proposed architectures explore this trade-off.* Let $\mu$ be a block expansion factor, defined as below:

$$\mu = \begin{cases} p \text{ for } p - LiTiChain \\ s \text{ for } s - LiTiChain \end{cases}$$

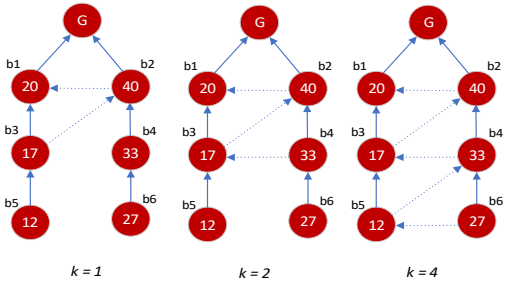

Fig. 5. Increase in number of AOG edges in LiTiChain with $K$

In LiTiChain, $\mu = 1$ for all the blocks. If $\mu = p$ or $s$ for at least a block in the chain, it is respectively called as *p-LiTiChain* or *s-LiTiChain*. The details are as follows.

*A. s-LiTiChain*

As discussed above, reducing blocksize is equal to splitting regular block of a LiTiChain into multiple sub-blocks, together known as *s-block*. Let *s* be the number of sub-blocks. As shown in Fig. 6, replacing a regular block with *s-block* increases the number of EOG edges and the height introduced by a block from 1 to *s*, which has the potential to reduce the number of AOG edges and overall retention cost of the chain for a given $K$. In Fig. 6, for $\mu = s = 2$, when $K$ is varied as 1, 2 and 4, overall retention cost of the chain increases as 0, 20 and 36 i.e. compared to the LiTiChain discussed in Fig. 5, retention cost is reduced for the same $K$. A LiTiChain that employs at least one *s-block* is known as *s-LiTiChain*. In this paper, to simplify the process of building *s-LiTiChain*, it is assumed that *s* takes only two values ($s = 1$ and another fixed value $s > 1$). In other words, *s-LiTiChain* can only have two types of blocks i.e. a regular block ($s = 1$) as in LiTiChain and an *s-block* ($s > 1$).

*B. p-LiTiChain*

Let us assume that for some reason, it is not a best option, or it is inconvenient to split a block into multiple sub-blocks. In this case, *s-LiTiChain* cannot be implemented and hence as an alternative, *p-LiTiChain* architecture is presented. In bitcoin, the number of transactions in a block are in the range of 1000-2000 and blocksize is around 1 MB. Let's assume a *Lightweight Block* (LWB) with few empty transactions (ex: 10) and size of around 10 KB i.e. approximately 100 times lighter than a regular block. As shown in Fig. 6, a *p-block* contains a regular block as in LiTiChain and *p-1* LWBs. Hence, if a regular block in a LiTiChain is replaced with a *p-block*, the number of EOG edges and height contribution of a block increases from 1 to *p*, which reduces the number of AOG edges and the overall retention cost of the chain. In Fig. 6, for $\mu = p = 2$, when $K$ is varied as 1, 2 and 4, retention cost varies as 0, 20 and 36 i.e. compared to the LiTiChain in Fig. 5, retention cost is reduced for a given $K$. A LiTiChain that contains at least one *p-block* is known as *p-LiTiChain*. In this paper, to reduce the complexity of *p-LiTiChain*, it is assumed that *p* takes only two values ($p = 1$ and another fixed value $p > 1$). In other words, a *p-LiTiChain* can only have two types of blocks i.e. a regular block ($p = 1$) as in LiTiChain and a *p-block* (with fixed $p > 1$).

The process of constructing *s-LiTiChain* and *p-LiTiChain* i.e. inserting and deleting blocks is as follows:

a) When the *blockheight* of the parent of a newly added block is less than or equal to $K$, the new block added to the chain

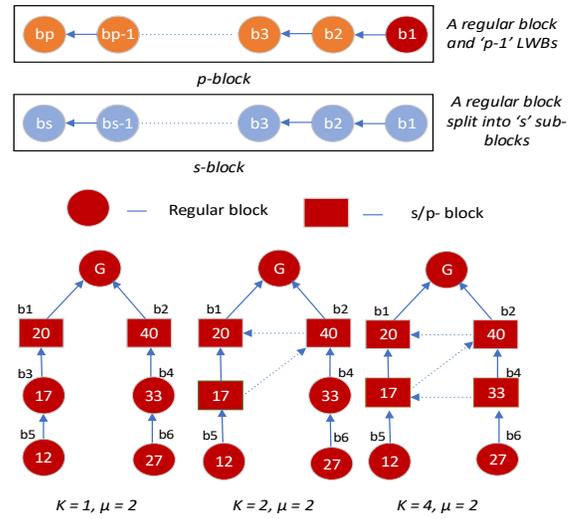

Fig. 6. Construction of *p-LiTiChain* and *s-LiTiChain*

is an *s/p-block* and an AOG edge is connected between block $b_i$ and $b_{i-1}$.

b) When the blockheight of the parent is greater than $K$, a regular block is inserted without an AOG edge.

For a given number of AOG edges, both *s-LiTiChain* and *p-LiTiChain* have the potential to increase the average *blockheight* of a LiTiChain, respectively by at least *s* and *p*. In other words, for the same number of AOG edges, both *s-LiTiChain* and *p-LiTiChain* offer better security compared to LiTiChain i.e. the effort required by an attacker to undo the chain increases compared to LiTiChain. In *p-LiTiChain*, even though the size of an LWB is assumed to be negligible i.e. 100 times smaller than a regular block, for higher values of *p*, the overhead due to LWBs is considerable i.e. for $\mu = p = 10$, LWB overhead is around 100 KB. Hence, for higher values of $\mu$, *s-LiTiChain* can be preferred over *p-LiTiChain*. On the other hand, as shown in Fig. 6, *p-LiTiChain* simplifies the blockchain design as it offers the same level of security as *s-LiTiChain* without varying the blocksize.

## V. PERFORMANCE EVALUATION

In this section, performance of *p-LiTiChain* and *s-LiTiChain* are evaluated using simulations. *Total Retention cost*, *average* and *maximum blockheight*, *total number of Nonce computations* are used as the performance metrics.

*A. Performance Metrics*

As mentioned earlier, $e_i$ is the expiration time of a block. Let $d_i$ be the deletion time of a block and $N$ denote the total number of blocks. Total Retention Cost ($\delta_K$) of all the blocks for a given value of blockheight threshold $K$ is given as follows:

$$\delta_K = \sum_{i=0}^{N}(d_i - e_i)$$

The *average blockheight H* and the *maximum blockheight M* are computed at the time of creating a new block and deleting an expired block. $H$ is obtained by taking average of blockheight over all the blocks that are alive in the chain.

Similarly, *M* is the maximum of the blockheights of all the blocks that are not expired in the chain. Let $\bar{H}$ and $\bar{M}$ denote the time averages of H and M during the lifetime of a blockchain. $\bar{H}$ and $\bar{M}$ are used as the performance metrics. The number of Nonce computations ($\epsilon$) of *p-LiTiChain* and *s-LiTiChain* are generally higher compared to LiTiChain. Hence, number of Nonce computations ($\epsilon_K$) for a given K is used as a performance metric.

*B. Simulation Setup*

In [21], realistic IoT data published by the New York City Taxi and Limousine Commission (TLC) with trip record data for yellow taxis is analyzed. The trip duration is considered as the lifetime of the transactions. Based on the simulations, it was concluded that, if the lifetime of blocks in a blockchain has bimodal distributions with small and large lifetime values, it would result in worst retention cost. The reason is that, compared to unimodal distributions, in bimodal distributions the short lifetime blocks suffer relatively more due to the time held back by the long lifetime blocks. For the purpose of simulations, in order to generate the lifetime data with a bimodal distribution, similar to [21], lifetime data is sampled from Z which is a mixture of the following two Gaussian distributions:

$$Z = \begin{cases} Z_1 & w.p. \ 0.5 \\ Z_2 & w.p. \ 0.5 \end{cases}$$

where $Z_1 \sim N(300, 110^2)$ and $Z_2 \sim N(1200, 110^2)$. Around 10000 lifetime data are sampled from the above distribution. The value of $\mu$ is varied between 1 and 500 and the values of *K* considered are 10, 50, 100, 300, 400 and 500.

*C. Simulation Results*

MATLAB is used to evaluate the performance of proposed architectures. As discussed above, in LiTiChain, as *K* increases, the number of AOG edges increases and hence $\delta_K$ increases. Fig. 7 shows the variation of relative $\delta_K$ w.r.t $\delta_{10}$ for different values of *K*. As expected, it is obvious from the simulations that $\delta_K$ increases with *K* in LiTiChain. Fig. 8 shows the variation in $\bar{H}$ and $\bar{M}$ for different values of *K* in LiTiChain. As expected, in LiTiChain, both $\bar{H}$ and $\bar{M}$ increase as the value of *K* is increased. But, as shown in Fig. 7, $\delta_K$ i.e. the additional storage cost incurred by the LiTiChain increases with increase in *K*. The proposed architectures aim at increasing $\bar{H}$ and $\bar{M}$ while reducing $\delta_K$.

To study the performance of *p-LiTiChain* and *s-LiTiChain*, two values of *K*, a lower value i.e. *K = 50* and a higher value i.e. *K = 500* are considered. Fig. 9 shows the variation in $\bar{H}$ and $\bar{M}$ w.r.t $\mu$ for *K = 50* and *500*. As the value of $\mu$ increases, $\bar{H}$ and $\bar{M}$ increase for both the values of *K*. For *K = 50*, $\mu \geq 10$ results in an average blockheight of $\bar{H} > 50$. Hence, $\mu \geq 10$ offers the required security as per the *longest chain rule* in the average sense. Fig. 10 shows the variation in relative $\delta_K$ w.r.t increase in $\mu$. As expected, retention cost $\delta_{50}$ decreases with increase in $\mu$ and it is zero when $\mu > K$ i.e. *when the value of block expansion factor is greater than the blockheight threshold, it is not required to extend the lifetime of the expired blocks and hence the additional storage costs are zero*. Hence, in Fig. 10, for *K = 50*, $\mu = 60$ results in zero retention cost.

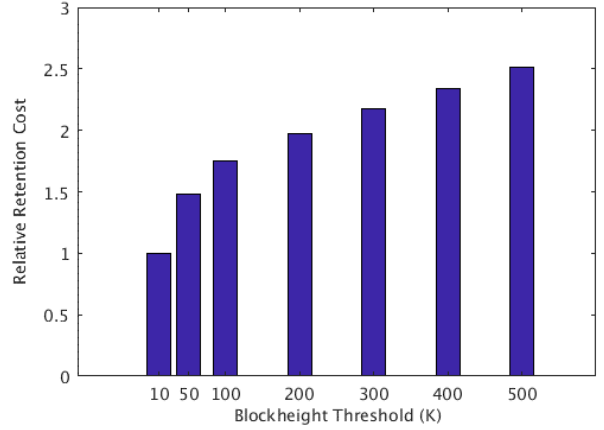

Fig. 7. Increase in Retention cost w.r.t Blockheight Threshold K in LiTiChain

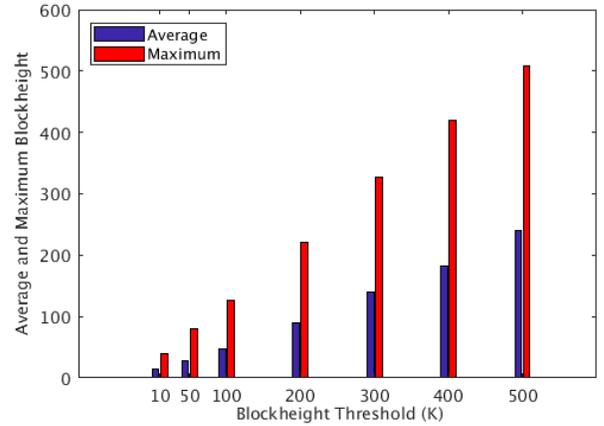

Fig. 8. Average and Maximum Blockheight for different values of *K* in LiTiChain

The lower retention cost and higher average blockheight $\bar{H}$ are achieved at the expense of increase in the total number of *Nonce Computations* ($\epsilon$). In other words, *the reduction in storage cost and improved security are obtained at the expense of higher computational costs*. Fig. 11 presents the relative increase in $\epsilon$ w.r.t $\mu$ for *K = 50* and *500*. As mentioned above, for *K = 50*, according to Fig. 9, $\mu \geq 10$ has the potential to offer the required security for the chain and as per Fig. 10, $\mu = 60$ results in zero retention cost. Also, as shown in Fig. 11, $\epsilon_{50}$ i.e. the total number of Nonce computations for $\mu = 60$ are around 6.14 times compared to $\mu = 1$. Hence, for *K = 50*, $\mu = 60$ offers the best performance in terms of storage cost, security and computational cost. On the other hand, as shown in Fig. 9, for *K = 500*, $\bar{H}$ is greater than *K* for $\mu = 60$. But, according to Fig. 10, $\delta_{500}$ decreases with $\mu$ but it does not reach zero as the value $\mu$ is not high enough. Similar to *K = 50*, there exists a value of $\mu$ for every *K* that offers optimal performance in terms of storage cost, security and computational cost.

The difference in the retention cost savings offered by *p-LiTiChain* and *s-LiTiChain* is as shown in Fig. 10. For a given *K*, the retention cost is same for both *p-LiTiChain* and *s-LiTiChain* when $\mu = 1$ for all the blocks. But, for $\mu \geq 1$ and $\mu \leq K$, as it employs LWBs, *p-LiTiChain* always results in higher retention cost compared to *s-LiTiChain*. For $\mu \geq K$, retention cost is zero for both *p-LiTiChain* and *s-LiTiChain*. Also, unlike *s-LiTiChain*, where the $\delta_K$ always decreases with increase in $\mu$, after a certain threshold value of $\mu$, the retention cost of *p-LiTiChain* tends to increase rather than decreasing. The reason for this is that, as explained in section IV, the LWB overhead becomes significant for higher values of $\mu$. Hence, as shown in Fig. 10, for $K = 500$ and $\mu = 100$, the retention cost of *p-LiTiChain* approximately doubles when compared to *s-LiTiChain*. As mentioned earlier in section IV, the size of LWBs is assumed to be 100 times lighter than regular blocks.

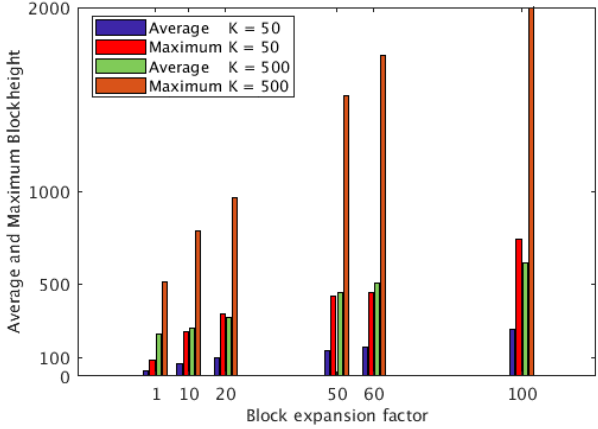

Fig. 9. Average and Maximum Blockheight w.r.t Block expansion factor $\mu$ for $K = 50$ and $500$

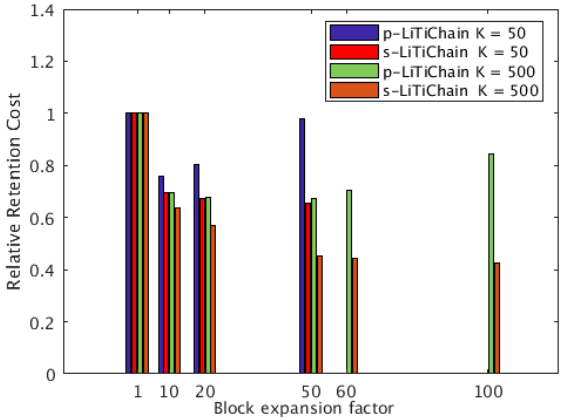

Fig. 10. Variation in Relative Retention Cost w.r.t Block expansion factor $\mu$ for K = 50 and 500 for *p-LiTiChain* and *s-LiTiChain*

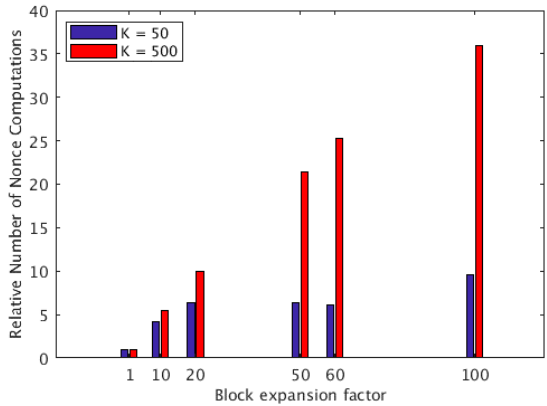

Fig. 11. Increase in Relative Number of Nonce Computations w.r.t Block expansion factor $\mu$ for K = 50 and K = 500

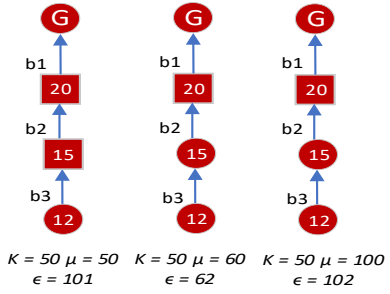

Fig. 12. Illustration of difference in number of *Nonce Computations* w.r.t $\mu$

Fig. 11 highlights another important point i.e. for $K = 50$, the total number of *Nonce* Computations decreases when $\mu$ increases from 50 to 60 and then again increases for $\mu = 100$. The reason is illustrated by Fig. 12. When $\mu = 50$, blockheight of $b_1$ is not greater than the threshold $K = 50$. Hence, an s/p-block is used for $b_2$. Therefore, the total number of *Nonce computations* $\epsilon = 101$. On the other hand, when $\mu = 60$ and $100$, blockheight of $b_1$ is greater than the threshold $K = 50$. Hence, a regular block is used for $b_2$ and the total number of *Nonce computations* are 62 and 102, respectively for $\mu = 60$ and $\mu = 100$.

## VI. CONCLUSION

This paper presented two new blockchain architectures i.e. *p-LiTiChain* and *s-LiTiChain* to design a blockchain of finite-lifetime blocks with applications to *Edge-IoT*. These two blockchain architectures are variants of LiTiChain published in the literature. With extensive simulations, it was shown that the proposed architectures have the potential to reduce storage cost and offer better security when compared to LiTiChain at the expense of computational costs. In conclusion, unlike the LiTiChain architecture presented in the literature, *p-LiTiChain* and *s-LiTiChain* offer a tradeoff between storage cost, security and computational cost which is worth exploring when designing blockchain for finite-lifetime data applications.